\documentclass[useAMS]{mn2e}
\usepackage{graphicx}
\usepackage{epsfig}
\usepackage{amssymb}
\usepackage{lscape}
\usepackage{ulem}
\usepackage{txfonts}

\def\aq{ALMaQUEST}

\def\sigh2{$\Sigma_{\rm H_2}$}

\def\sigstar{$\Sigma_{\star}$}

\def\c2s{C\,{\sc ii}$^{\star}$}

\def\fgas{$f_{\rm H_2}$}

\title[AGN in EDGE-CALIFA] {The EDGE-CALIFA Survey: Central molecular gas depletion in AGN host galaxies - a smoking gun for quenching?}

\author[Ellison et al.] {Sara L. Ellison$^1$, Tony Wong$^2$, Sebastian F. S\'{a}nchez$^3$,  Dario Colombo$^4$,  Alberto Bolatto$^5$,
  \newauthor Jorge Barrera-Ballesteros$^3$, Rub\'en Garc\'ia-Benito$^6$, Veselina Kalinova$^4$,  Yufeng Luo$^2$, \newauthor
  Monica Rubio$^7$, Stuart N. Vogel$^5$ \\
$^1$ Department of Physics \& Astronomy, University of Victoria, Finnerty Road, Victoria, British Columbia, 
  V8P 1A1, Canada \\
$^2$  Department of Astronomy, University of Illinois, Urbana, IL 61801, USA\\
  $^3$  Instituto de Astronomia, Universidad Nacional Autonoma de Mexico, A. P. 70-264, C.P. 04510, Mexico, D.F., Mexico\\
  $^4$ Max-Planck-Institut f\"ur Radioastronomie, Auf dem H\"ugel 69, 53121 Bonn, Germany\\
  $^5$ Department of Astronomy, University of Maryland, College Park, MD 20742, USA\\
  $^6$ Instituto de Astrof\'isica de Andaluc\'ia – CSIC, Apdo. 3004, 18080 Granada, Spain\\
  $^7$ Departamento de Astronomia, Universidad de Chile, Casilla 36-D, Santiago, Chile
}

\begin{document}

\maketitle

\begin{abstract}
Feedback from an active galactic nucleus (AGN) is often implicated as a mechanism that leads to the quenching of galactic star formation.  However, AGN-driven quenching is challenging to reconcile with observations that AGN hosts tend to harbour equal (or even excess) amounts of gas compared with inactive galaxies of similar stellar mass.  In this paper, we investigate whether AGN feedback happens on sub-galactic (kpc) scales, an effect that might be difficult to detect with global gas measurements.  Using kpc-scale measurements of molecular gas (\sigh2) and stellar mass (\sigstar) surface densities taken from the EDGE-CALIFA survey, we show that the gas fractions of central AGN regions are typically a factor of $\sim$ 2 lower than in star-forming regions.  Based on four galaxies with the best spaxel statistics, the difference between AGN and star-forming gas fractions is seen even within a given galaxy, indicating that AGN feedback is able to deplete the molecular gas reservoir in the central few kpc.   
\end{abstract}


\section{Introduction}\label{intro_sec}

The issue of why galaxies cease to form stars is a long-standing question in astrophysics.  One of the prime suspects for quenching star formation is feedback from an active galactic nucleus (AGN).  The common occurence of jets and outflows driven by nuclear activity  (e.g. Harrison et al. 2014; Woo et al. 2016; Concas et al. 2017) offer a logical energy source for the heating or ejection of galactic gas (e.g. Zinger et al. 2020).   

Large galaxy surveys have provided abundant statistical evidence supporting the notion of widespread AGN-driven quenching.  The host galaxies of AGN selected in the optical and radio tend to have star formation rates (SFRs) that are lower than typical main sequence galaxies (e.g. Ellison et al. 2016; Smith et al. 2016; S\'{a}nchez et al. 2018; Lacerda et al. 2020).  Integral field unit (IFU) surveys of galaxies have revealed that galaxies that exhibit low SFRs tend to be preferentially suppressed in the inner regions, suggestive of a quenching mechanism that operates from the inside-out (Ellison et al. 2018; Medling et al. 2018; S\'{a}nchez et al. 2018;  Wang et al. 2019; Kalinova et al. 2021).  Galaxies with more massive/dominant bulges (e.g. Bluck et al. 2014, 2016, 2020; Omand et al. 2014; Teimoorinia et al. 2016), higher Sersic indices (e.g. Wuyts et al. 2012; Mendel et al. 2013; Ellison et al. 2021a) or those with higher central stellar mass concentrations (e.g. Cheung et al. 2012; Fang et al. 2013; Woo et al. 2015) are more likely to be quenched.  Bulge dominated galaxies also have lower star formation efficiencies (Colombo et al. 2018; Eales et al. 2019; Ellison et al 2021a), although radial profiles show considerable variation (Utomo et al. 2017).  Given the close connection between bulge and black hole growth, correlations between structure and star formation (i.e. the morphology-quiescence relation, Woo \& Ellison 2019) have often been used to imply AGN-driven quenching.  The connection between nuclear activity and quenching is further supported by studies using direct measurements of black hole masses, in which quiescence is preferentially associated with higher M$_{BH}$ (Terrazas et al. 2016, 2017).  The successful reproduction of observed galaxy properties in hydrodynamical simulations is also achieved by implementing AGN feedback (e.g. Terrazas et al. 2020).

If AGN are responsible for quenching, it is natural to expect that the gas reservoirs of their host galaxies will be impacted.  However, numerous observational studies of low redshift\footnote{In contrast, at high redshift, several studies \textit{have} reported low molecular gas fractions see e.g. Circosta et al. (2021) and references therein for a discussion on possible biases and limitations in high redshift work.} galaxies have shown that the atomic (e.g. Ho et al 2008; Fabello et al. 2011; Ellison et al. 2019), molecular (Maiolino et al. 1997;  Saintonge et al. 2017; Shangguan et al. 2020; Koss et al. 2021) and total (Vito et al. 2014; Shangguan, Ho \& Xie 2018; Shangguan \& Ho 2019) gas content of AGN hosts is consistent with (or even higher than) galaxies of similar mass and morphology that do not harbour an AGN.  This is true even for the most powerful AGN and those known to exhibit powerful outflows (e.g. Jarvis et al. 2020).  A minority of studies \textit{have} reported lower molecular gas fractions in AGN hosts, e.g. S\'{a}nchez et al. (2018) and Lacerda et al. (2020), but these have been based on indirect estimates of the gas content that may be less accurate for AGN hosts (Concas \& Popesso 2019).

In the work presented here, we explore the possibility that the lack of an obvious impact of the AGN on a galaxy's gas reservoir is due to the scale over which feedback is felt.  Since many of the outflows that have been identified in AGN hosts exist on scales of a few kpc (Karouzos et al. 2016; Kang \& Woo 2018) and the suppression of star formation in quenched galaxies is observed on similar scales (Ellison et al. 2018), studies of the global gas reservoirs may be insensitive to more local signatures of AGN feedback, if the majority of the galactic gas supply is unaffected.  Instead, AGN feedback effects may be limited to the central regions (e.g. Fluetsch et al. 2019).  In present study, we use kpc-scale maps of molecular gas (\sigh2) and stellar mass (\sigstar) surface densities data from the Extragalactic Database for Galaxy Evolution-Calar Alto Legacy Integral Field Area (EDGE-CALIFA) survey to investigate whether the central regions of optically selected (Seyfert) AGN exhibit low molecular gas fractions compared with star forming regions.

\section{Data}\label{data_sec}

The data used in the work presented here combine optical IFU maps from the CALIFA survey (S\'{a}nchez et al. 2012) with CO(1-0) observations obtained from the EDGE survey (Bolatto et al. 2017) for 126 galaxies.  We hereafter refer to this combined dataset as the EDGE-CALIFA survey.    In order to generate maps on the same spatial scale, both the CALIFA and EDGE datasets are smoothed to 7 arcsec, which is the resolution of the poorest CO dataset.  The spaxel size is 1 arcsec, and therefore the point spread function is heavily over-sampled.  We have repeated the analysis presented here with both the native sampling, as well as a subset of approximately Nyquist sampled spaxels spaced by 3 arcseconds, and draw the same qualitative conclusions.

Emission line fluxes and stellar mass surface densities are extracted from the degraded datacubes using PIPE3D (S\'{a}nchez et al. 2016).  We adopt a Salpeter initial mass function and a constant conversion factor $\alpha_{CO}$=4.3 M$_{\odot}$ pc$^{-2}$ (K km s$^{-1}$)$^{-1}$ (e.g. Bolatto et al. 2013). Emission line fluxes are corrected using the Balmer decrement, assuming an intrinsic H$\alpha$/H$\beta$ = 2.86 and a Milky Way extinction curve (Cardelli, Clayton \& Mathis 1989) with $R_v$=3.1.  All surface densities are inclination corrected using axial ratios from the LEDA database (Makarov et al. 2014).

For the work presented here, we require that CO is detected with a S/N$>$2.  We repeated the analysis with stricter cuts of S/N$>$ 3, 4 or 5 and found no qualitative difference in our results.  From these CO-detected spaxels we further classify spaxels according to their optical emission line properties:

\begin{itemize}

\item Star-forming spaxels must have a S/N$>$2 in H$\alpha$, H$\beta$, [NII]$\lambda$6584 and [OIII]$\lambda$5007, must be located below the Kauffmann et al. (2003) demarcation and have an H$\alpha$ equivalent width (EW) $>$ 6 \AA.

\item AGN spaxels must have a S/N$>$2 in H$\alpha$, H$\beta$, [NII]$\lambda$6584 and [OIII]$\lambda$5007, be above the Kewley et al. (2001) demarcation, have an H$\alpha$ EW $>$ 6 \AA\ and be centrally located (within 1$R_e$, calculated by growth curve photometry performed using elliptical apertures on the SDSS DR7 $r$-band images; Walcher et al. 2014).  We also require that the AGN spaxels do not belong to any of the Type I AGN in CALIFA identified in Lacerda et al. (2020), as these broad emission line galaxies are not well fit by PIPE3D and the derived physical quantities are therefore likely to be erroneous.

\item Retired spaxels must have an H$\alpha$ equivalent width (EW) $<$ 3 \AA\ (e.g. Stasinska et al. 2008; Cid-Fernandes et al. 2011).

\end{itemize}

Finally, we remove (visually identified) galaxy mergers from the sample in case galaxy-galaxy interactions significantly impact the central molecular gas reservoir (e.g. Ueda et al. 2014), or generate shocks that could falsely masquerade as AGN.  These factors could complicate our interpretation of the results and are beyond the scope of this Letter.

\section{Results}\label{results_sec}

\begin{figure}
	\includegraphics[width=9cm]{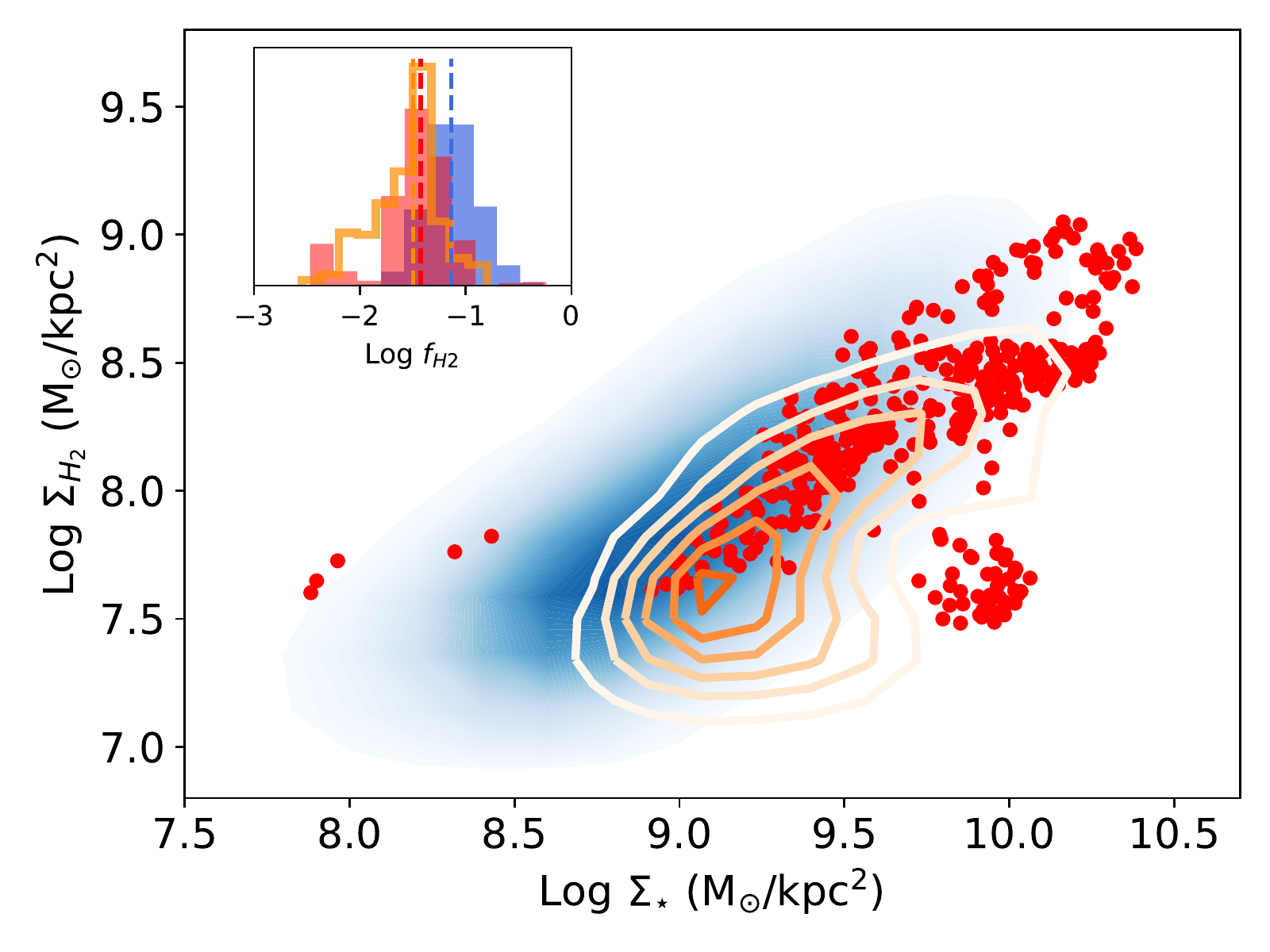}
        \caption{The resolved molecular gas main sequence for centrally located ($R< R_e$) 1 arcsecond spaxels classified as star forming (blue shading), retired (orange contours) and AGN (red points) in the EDGE-CALIFA survey.   The inset histogram shows the (normalized) distribution of molecular gas fractions of the three populations. The (logarithmic) median gas fractions (shown by vertical dashed lines) of the star forming, retired and AGN populations are $-1.14$, $-1.49$ and $-1.42$ dex, respectively.  Both the retired and AGN spaxels therefore have lower \sigh2\ at fixed \sigstar, indicating gas fractions that are typically a factor of $\sim$ 2 lower than the star-forming population.}
    \label{mgms_ensemble}
\end{figure}

The resolved molecular gas main sequence (rMGMS) is the tight correlation between \sigstar\ and \sigh2\ observed on kpc-scales for star-forming spaxels (e.g. Lin et al. 2019; Barrera-Ballesteros et al. 2020; Ellison et al. 2021a; S\'{a}nchez et al. 2021).  Using the ALMA MaNGA QUEnching and STar formation (ALMaQUEST) survey (Lin et al. 2020), Ellison et al. (2021b) have recently shown that retired spaxels form a distinct rMGMS that is offset to low \sigh2\ at fixed \sigstar, indicating that galactic regions that are on the road to quenching have lower gas molecular fractions (defined here as \fgas\ = \sigh2 / \sigstar) than the star forming regions.   In Fig. \ref{mgms_ensemble} we show the rMGMS for star-forming (blue-shaded region) and retired (orange contours) spaxels in the EDGE-CALIFA sample (see also S\'{a}nchez et al. 2021).  Only spaxels within 1 R$_e$ are shown, in order that we might later compare with the centrally located AGN spaxels, which have the same radial cut.  These centrally located spaxels are drawn from 90 (star-forming) and 21 (retired) unique galaxies.  As was found by Ellison et al. (2021b) for the ALMaQUEST sample, the retired regions in the EDGE-CALIFA sample have low \sigh2\ at fixed \sigstar.  The gas fractions of the two spaxel populations are shown in the inset histogram of Fig. \ref{mgms_ensemble}.  Although qualitatively similar to the results of Ellison et al. (2021b), the difference in gas fraction between the star-forming and retired spaxels is smaller for EDGE-CALIFA (a factor of $\sim2.5$) than for ALMaQUEST (a factor of $\sim 5$).  Both samples have similar median gas fractions for the star-forming populations (about 10 per cent, a value that is little affected by the $R<R_e$ cut used in Fig. \ref{mgms_ensemble}), but the median retired gas fraction of retired spaxels is significantly lower in ALMaQUEST (log\fgas\ = $-1.9$) than EDGE-CALIFA  (log\fgas\ $= -1.5$).  The difference between star-forming and retired gas fractions in the two samples may be due to the deeper CO images in \aq\ compared with EDGE.

Also plotted in Fig. \ref{mgms_ensemble} as red points are the spaxels identified as ionized by a central AGN; the AGN spaxels orignate from five distinct galaxies.   AGN spaxels tend to be located at higher \sigstar\ than star-forming spaxels due to their central locations.  Nonetheless, at fixed \sigstar, AGN spaxels have lower \sigh2\ than the star-forming spaxels.  Although Fig. \ref{mgms_ensemble} only shows spaxels detected in CO (the CO detection fractions in these central regions are 67\%, 53\% and 30\% for AGN, star-forming and retired spaxels, respectively), the AGN (and retired) spaxels clearly do not reach gas fractions as high as the star-forming population. 

Ellison et al. (2021b) showed that the retired spaxels in the ALMaQUEST sample are preferentially located in the centres of galaxies (see also Belfiore et al. 2017, 2018; S\'{a}nchez et al. 2018), a feature which further supports inside-out quenching.  The common central locations of retired and AGN galaxies (albeit in different galaxies) is possible evidence for a causal link between the AGN episode and the cessation of star formation.  However, we note that the retired spaxels plotted in Fig. \ref{mgms_ensemble} are drawn from different galaxies than the AGN spaxels.  Moreover, deeper CO observations would be required to more completely assess the similarity of gas fractions in the AGN and retired populations.

Fig. \ref{mgms_ensemble} demonstrates that the AGN spaxels in the EDGE-CALIFA sample have lower gas fractions than star-forming spaxels, on average.  In order to investigate the internal impact of the AGN, we next investigate the rMGMS on a galaxy-by-galaxy basis.  Following Ellison et al. (2021b), we select EDGE-CALIFA galaxies that contain at least 40 star-forming and at least 40 central AGN spaxels, excluding mergers and Type I AGN.  Although this choice of spaxel count is somewhat arbitrary, for EDGE-CALIFA a grouping of 40 spaxels is approximately one independent beam for the poorest resolution data (and our results are not significantly affected by other reasonable choices).  Four galaxies are thus selected:  NGC 2410, NGC 2639, NGC 6394 and IC 2247, all of which have previously been identified as Type II AGN by Lacerda et al. (2020), and are also identified as strong AGN by Kalinova et al. (2021).   The fifth galaxy in the EDGE-CALIFA sample that contributes AGN spaxels to Fig. \ref{mgms_ensemble}, NGC 7738, is excluded from the galaxy-by-galaxy analysis due to a lack of star-forming spaxels for comparison.  

\begin{figure}
	\includegraphics[width=9cm]{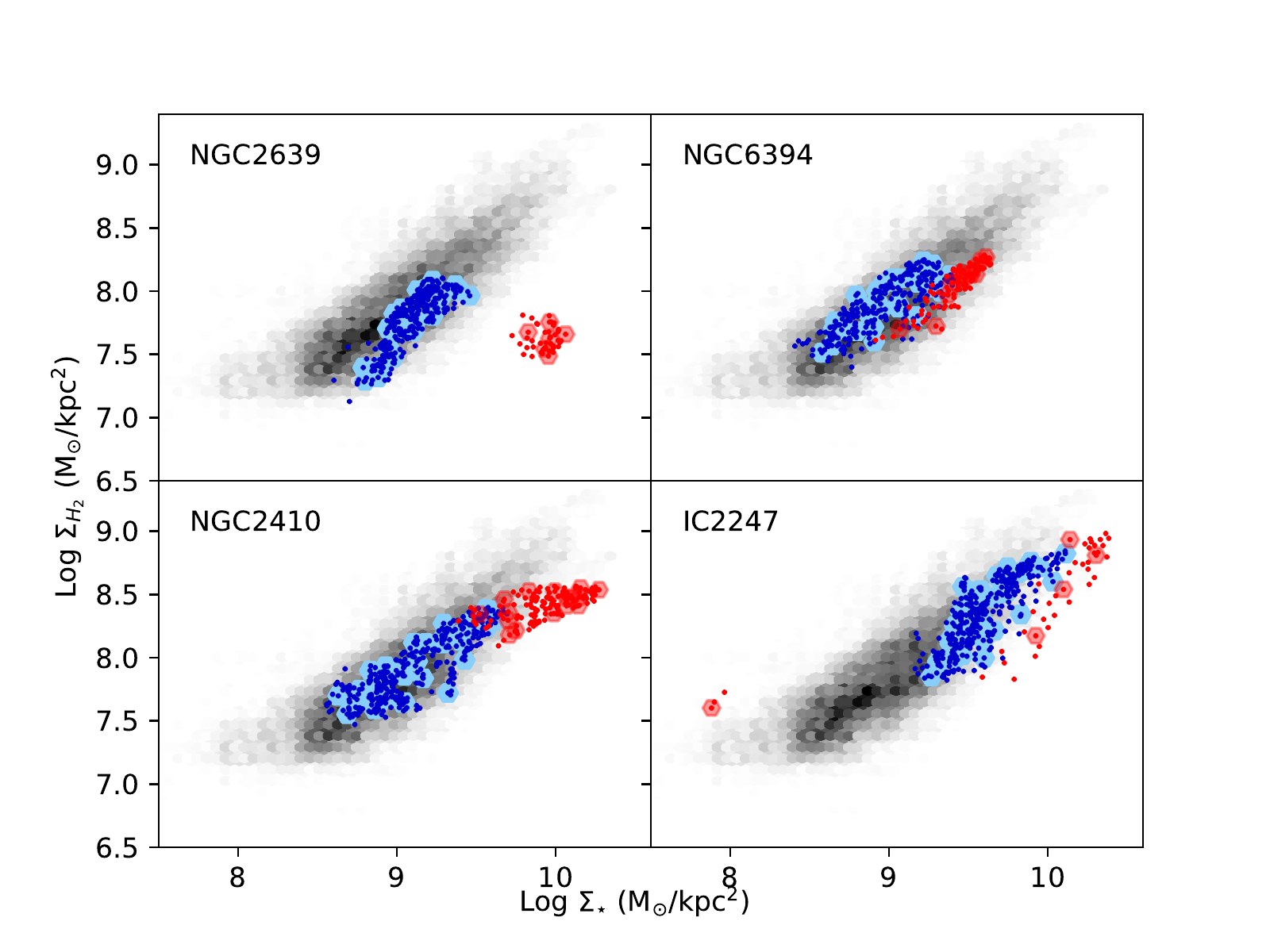}
        \caption{The rMGMS for four AGN host galaxies in our sample.  Star-forming spaxels are shown in blue and AGN spaxels are shown in red.  Small points show the locations of 1 arcsecond spaxels and large pale hexagons show the Nyquist sampled pixels. The AGN spaxels are offset from the star-forming spaxels in the same galaxy.  The greyscale background is the same in each panel and shows the ensemble of all star-forming spaxels with $R<R_e$ in EDGE-CALIFA (i.e. the spaxels contributing to the blue shaded region in Fig \ref{mgms_ensemble}), for reference. }
    \label{mgms_galaxy}
\end{figure}

In Fig. \ref{mgms_galaxy} we show the rMGMS on a galaxy-by-galaxy basis for each of the four AGN galaxies listed above.  In each panel the greyscale background shows the rMGMS for the ensemble of EDGE-CALIFA star-forming spaxels with $R<R_e$ (i.e. the same sample as shown in blue in Fig. \ref{mgms_ensemble}) for visual reference.  The blue and red points in each panel show the rMGMS for the star-forming and AGN spaxels in a given galaxy (none of the four galaxies has a significant number of retired spaxels).  Small points show the locations of 1 arcsecond spaxels and large pale hexagons show the Nyquist sampled pixels.  The fraction of star-forming (blue) and AGN (red) spaxels detected in CO in each galaxy is also indicated in each panel.  For 3/4 of the galaxies CO is detected in $\sim$ 100 per cent of the AGN spaxels, indicating that we have a fairly complete picture of the gas fractions in these central regions.  Fig. \ref{mgms_galaxy} shows that the star-forming spaxels in a given galaxy form a tight rMGMS, although the locus of this position varies from galaxy-to-galaxy (see also Ellison et al. 2021a for a larger sample of non-AGN \aq\ galaxies).  In any given galaxy, the AGN-dominated spaxels are offset from its star-forming spaxels.  This is due, at least in part, to the fact that the AGN are centrally concentrated and are hence located at higher values of \sigstar\ than star-forming spaxels in the same galaxy.  However, in all four galaxies, the AGN are offset from an extrapolation of the rMGMS of its star-forming spaxels, indicating that this is not simply a radius effect (as also argued above for the ensemble sample in Fig. \ref{mgms_ensemble}). 

\begin{figure}
	\includegraphics[width=9cm]{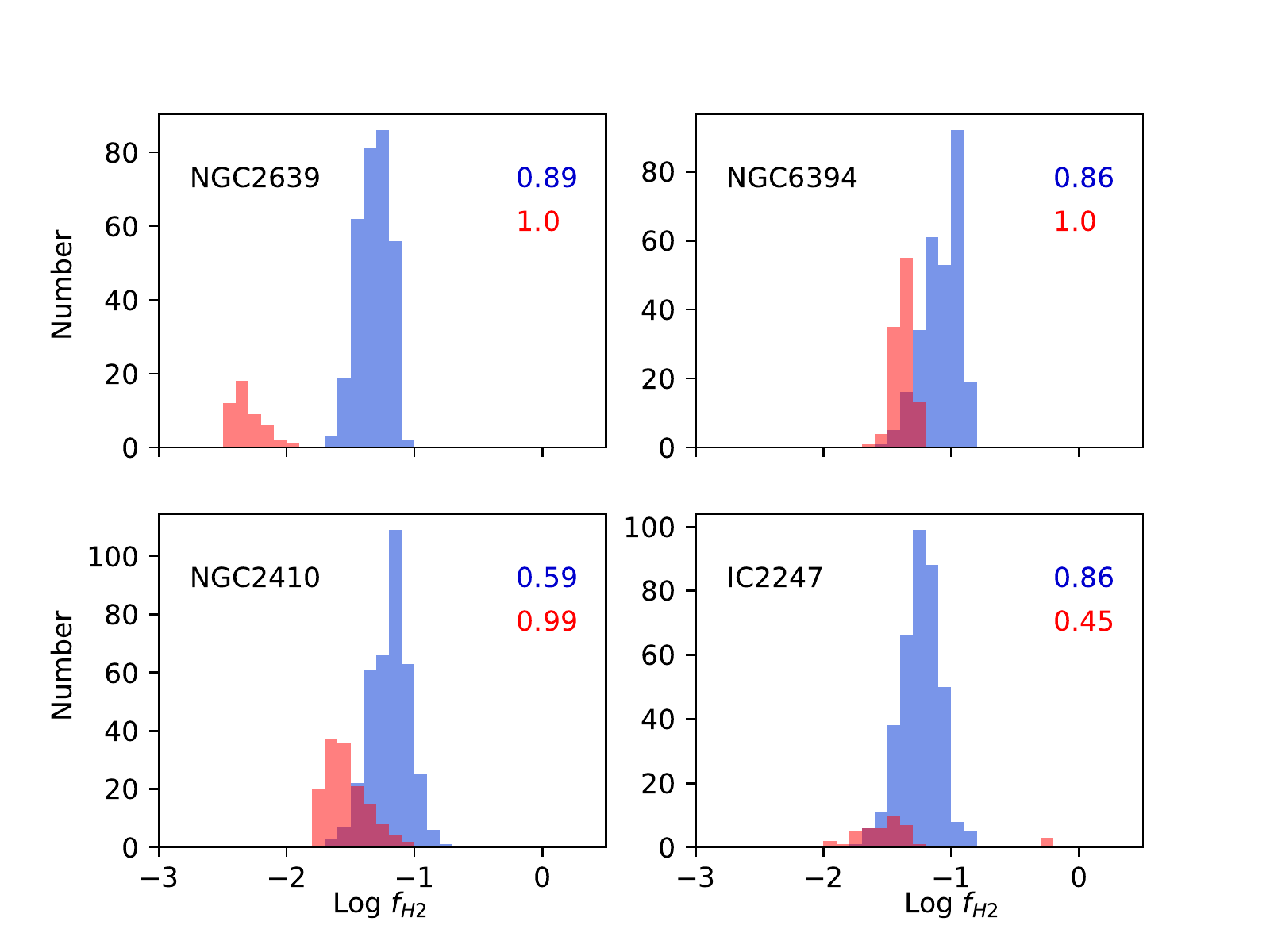}
        \caption{Gas fraction distributions for four AGN host galaxies in our sample.  The distribution for star-forming spaxels is shown in blue and for AGN spaxels in red.  The fraction of star-forming/AGN spaxels detected in CO (S/N$>$2) in the top right.  Panels appear in the same order as Fig. \ref{mgms_galaxy}.  AGN spaxels have lower molecular gas fractions by up to an order of magnitude compared with star-forming spaxels in the same galaxy.}
    \label{fgas}
\end{figure}

Further evidence for an offset between the rMGMS of star-forming and AGN spaxels is presented in Fig. \ref{fgas}, in which we plot the distribution of molecular gas fraction on a galaxy-by-galaxy basis.  The legend in the top right of each panel indicates the fraction of CO detected spaxels amongst the star-forming (blue) and AGN (red) populations in that galaxy.  The gas fraction distribution of AGN spaxels is offset to lower values than the star-forming spaxels in all four galaxies, typically by a factor of a few, but by more than an order of magnitude in NGC 2639.  It is also clear from Fig. \ref{fgas} that the star-forming spaxels in each of the four galaxies significantly outnumber the AGN spaxels.  Therefore, by studying the gas on these sub-galactic scales, and distinguishing the central AGN dominated regions, distinct molecular gas fractions are clearly present.  Our results support the hypothesis that previous studies that have found normal \textit{global} gas fractions in AGN may have been insensitive to the relatively compact regions of gas depletion.  Indeed, despite their low central gas fractions the integrated molecular gas masses in the AGN hosts of our sample are largely consistent with those of star-forming galaxies of the same stellar mass (Colombo et al., in prep).  

Although all four of the galaxies in Fig. \ref{fgas} have low molecular gas fractions in their central AGN regions compared with the extended star-forming regions, NGC 2639 is particularly poor in molecular gas.  We investigate whether this may be due to the relative luminosity of the AGN, which we infer from the luminosity of the [OIII] $\lambda$5007 emission line.  Most of the AGN spaxels in NGC 2639 have log L([OIII])$>$ 40 ergs s$^{-1}$, whereas the AGN spaxels in the other galaxies rarely have luminosities above this value, indicating that the level of molecular gas depletion may be governed by AGN luminosity.  Our results are consistent with previous findings of a dependence of both outflow rate and  depletion time on AGN luminosity (Cicone et al. 2014; Fluetsch et al. 2019).  NGC 2639 is also known to have a complex radio jet structure (e.g. Sebastian et al. 2019), and is the most bulge dominated galaxy in our sample.  Extending the work presented here to larger, more diverse AGN host galaxy samples is necessary to better understand the galaxy-to-galaxy impact of nuclear feedback.

\section{Summary and Discussion}

The main result of the work presented here is that the central (kpc-scale) regions of AGN host galaxies have lower gas fractions than star-forming regions.  There have been two previous (statistical) studies that have also specifically studied the gas content in the inner regions of AGN hosts.   S\'{a}nchez et al. (2018) inferred the total gas surface density, $\Sigma_{gas}$, from the $A_V$ calibration of Barrera-Ballesteros et al. (2020) and found preferentially less gas in the inner regions of AGN hosts in the MaNGA IFU survey, at a similar spatial resolution to the EDGE-CALIFA sample.  Although qualitatively consistent with our result, Concas \& Popesso (2019) caution the use of such $A_V$-based calibrations for AGN regions.  Indeed, we find that the AGN spaxels in our sample do not follow the same trend in $A_V$ - \sigh2\ as the star-forming spaxels.  Our work therefore represents an important follow-up of S\'{a}nchez et al. (2018), by using direct molecular gas measurements to complement the optical IFU data.

In a complementary work, Rosario et al. (2018) made single dish CO(2-1) measurements of the inner kpc regions of 17 low redshift X-ray selected AGN and compared these centrally measured molecular gas fractions to those of a sample of inactive control galaxies.  No difference was found in the central molecular gas fractions of the AGN and control samples, apparently at odds with the results presented here.  However, the central AGN gas fractions in both studies are consistent.  We therefore speculate that the apparently conflicting results of Rosario et al. (2018) and the work presented here may be due to the different comparison/control techniques of the two studies.  Our comparison is explicitly between AGN spaxels and star-forming spaxels, either in the ensemble sample (Fig. \ref{mgms_ensemble}) or within a given galaxy (Fig. \ref{mgms_galaxy}).  Rosario et al. (2018) compare central AGN gas fractions to a control sample matched in stellar mass and Hubble type.  As shown in Fig. \ref{mgms_ensemble}, the central `retired' regions of galaxies have similarly low gas fractions as the AGN spaxels (Ellison et al. 2021b and Fig. \ref{mgms_ensemble}). Therefore, if the centres of control galaxies are already `retired', there would be no detectable difference in their gas fractions compared with central AGN regions.

In the current work we have assumed a constant $\alpha_{CO}$=4.3 M$_{\odot}$ pc$^{-2}$ (K km s$^{-1}$)$^{-1}$ for all spaxels.    However, the central regions of galaxies can exhibit lower values of $\alpha_{CO}$ (e.g. Bolatto et al. 2013; Sandstrom et al. 2013).  A lower value of $\alpha_{CO}$ applied to the central AGN spaxels would result in even lower molecular gas fractions, accentuating our finding of centrally diminished \fgas.  X-rays, shocks and cosmic rays could potentially reduce the CO/H$_2$ ratio (i.e. increase $\alpha_{CO}$), which could in turn lead to a reduction in the observed amount of CO(1-0), e.g. Bisbas et al. (2017).  However, Shangguan et al. (2020) found no evidence for a distinct conversion factor in AGN host galaxies. We conclude that a deficit of molecular gas is the most straightforward explanation of our results.

The IllustrisTNG simulation has recently achieved success in reproducing observed trends in star formation and quenching (e.g. Terrazas et al. 2020; Nelson et al. 2021) through the implementation of a low accretion rate kinetic mode of AGN feedback (Weinberger et al. 2017;  Zinger et al. 2020).  The long-lasting imprint of AGN feedback on the large scale gas reservoirs (Zinger et al. 2020) means that it is the integrated black hole accretion (captured by the black hole mass), rather than the instantaneous accretion that determines whether or not a galaxy will be quenched in the long-term (Terrazas et al. 2020; Piotrowska et al. in prep).  Thus, the mechanism for quenching in IllustrisTNG might appear to contrast with our observations for a prompt signature of AGN feedback on the ISM in the observations.    However, the kinetic mode of AGN feedback implemented in Illustris TNG leads to the removal of gas that begins in the central regions (Nelson et al. 2019; Terrazas et al. 2020), resulting in a pattern of inside-out quenching (Nelson et al. 2021).  As emphasized by Zinger et al. (2020), the feedback in IllustrisTNG is therefore both ejective (removing the central gas reservoir) and preventative (heating the circumgalactic gas), leading to both prompt feedback in the inner regions and a long-term effect on suppressing star formation through the impact on the halo gas.  The observations presented here support the signature of the former of these effects, although it is unclear whether the current epoch of cold molecular gas depletion will lead to the long-term suppression of star formation throughout the AGN host galaxies.

\medskip

In summary, using kpc-scale maps of stellar mass and molecular gas from 126 galaxies in the EDGE-CALIFA sample, we have investigated the gas fractions in AGN host galaxies on the sub-galactic scale.   Our main conclusions are:

\begin{enumerate}
  
\item  In the ensemble of survey spaxels, central (kpc-scale) AGN regions have gas fractions that are typically a factor of $\sim$2 lower than regions photoionized by stars, suggesting that the AGN has (partially) depleted the central molecular gas reservoir (Fig. \ref{mgms_ensemble}).
  
\smallskip  

\item  Four galaxies are identified with at least 40 AGN and 40 star-forming spaxels in order to compare the molecular gas contents of these regions within a given galaxy.  The offset between the rMGMS of AGN and star-forming spaxels is seen in each of these individual galaxies (Fig. \ref{mgms_galaxy}).

  \smallskip

\item  In each of the four EDGE-CALIFA galaxies in which at least 40 AGN and 40 star-forming spaxels are detected in CO, the central AGN dominated region has a lower gas fraction (by up to an order of magnitude) than the extended star-forming region.  The greatest molecular gas depletion is observed at the highest AGN luminosity.  The dominance of star-forming spaxels means that global studies could potentially be insensitive to the effect of the central AGN on molecular gas fraction  (Fig. \ref{fgas}).
 
\end{enumerate}


\section*{Acknowledgements}

The authors acknowledge an NSERC Discovery Grant (SLE), NSF AST-1616199 (TW), support by the Deut\-sche For\-schungs\-ge\-mein\-schaft project number SFB956A (DC), NSF AST-1615960 (AB, SNV), PID2019-109067GB-I00, P18-FRJ-2595, SEV-2017-0709 (RGB), IA-100420, IN100519, CF19-39578, CB-285080, and FC-2016-01-1916 (SFS and JBB), Fondecyt grant No 1190684 (MR).  Support for CARMA construction was derived from the Gordon and Betty Moore Foundation, the Eileen and Kenneth Norris Foundation, the Caltech Associates, the states of California, Illinois, and Maryland, and the NSF. Funding for CARMA development and operations were supported by NSF and the CARMA partner universities.  We acknowledge the usage of the HyperLeda database (http://leda.univ-lyon1.fr).

\section*{Data Availability}

This article is based on publicly available data from the CALIFA (https://califa.caha.es) and EDGE (https://www.astro.umd.edu/EDGE/) surveys.

\end{document}